\documentclass{elsart}
\usepackage{graphicx}
\newcommand{\chha}{$\tilde{\chi}_1^{\pm}$}

\newcommand{\chna}{$\tilde{\chi}_1^{0}$}
\newcommand{\chnb}{$\tilde{\chi}_2^{0}$}

\newcommand{\g}{$\tilde{g}$}
\newcommand{\q}{$\tilde{q}$}
\newcommand{\etm}{$E_T^{miss}$}
\def\gsim{\mathrel{\rlap{\raise.4ex\hbox{$>$}}{\lower.6ex\hbox{$\sim$}}}}
\def\lsim{\mathrel{\rlap{\raise.4ex\hbox{$<$}} {\lower.6ex\hbox{$\sim$}}}}
\begin{document}

\begin{frontmatter}
\title{Search for the next-to-lightest neutralino}

\author{I.~Iashvili and A.~Kharchilava}

\address{Institute of Physics, Georgian Academy of Sciences,
6 Tamarashvili, 380077 Tbilisi, Georgia}

\begin{abstract}   

We study the inclusive production of the next-to-lightest 
neutralino \chnb, decaying directly, or via slepton, into two leptons
and the lightest neutralino. The dilepton invariant mass
spectrum in these decays has a characteristic sharp edge
near the kinematical upper limit. We propose to exploit this
feature as a search strategy for the \chnb, and thereby
for SUSY. The possibilities to determine neutralinos and
slepton masses are also discussed.

\hspace*{-4mm}{\it PACS:} 14.80.Ly, 13.85.Qk; {\it keywords:} SUSY,
neutralino, slepton, dilepton mass spectrum.

\end{abstract}
\end{frontmatter}

\section{Introduction}

The Standard Model (SM), despite its phenomenological 
successes, is most likely a low energy effective theory
of matter fermions interacting via gauge bosons. A good 
candidate for the new physics is the 
Supersymmetry (SUSY), which in its minimal version introduces
a scalar (fermion)
partner to each ordinary fermion (boson) with the same couplings.
If $R$-parity is conserved, SUSY particles are
produced in pairs and a stable Lightest Supersymmetric Particle (LSP) 
appears at the end of each sparticle decay chain. Weakly interacting
LSP escapes detection and masses of sparticles cannot be reconstructed
explicitly. Usually, one characterises the SUSY
signal significance by an excess of events over the SM
background expectation.

SUSY partners of gauge and Higgs bosons mix to form physical
states, charginos $\tilde{\chi}^{\pm}_{1,2}$,
and neutralinos, $\tilde{\chi}^0_{1,2,3,4}$.
Depending on a particular SUSY scenario the decays
\chnb $\rightarrow l^+ l^-$\chna \
and/or 
\chnb $\rightarrow l^{\pm} \tilde{l}^{\mp} \rightarrow l^+ l^-$\chna \ 
can take sizable branching ratios. The dilepton
invariant mass spectrum in these decays has a specific shape with the
sharp edge near the kinematical upper limit. 
In case of three-body (direct) decays \chnb $\rightarrow l^+ l^-$\chna \ 
the  position of edge is
\begin{equation}
M^{max}_{l^{+}l^{-}} = M_{\tilde{\chi}^{0}_{2}} - M_{\tilde{\chi}^{0}_{1}}
\end{equation}
and in case of two-body (cascade) decays it is 
\begin{equation}
M^{max}_{l^{+}l^{-}} = \frac{\sqrt{
(M^{2}_{\tilde{\chi}^{0}_{2}} - M^{2}_{\tilde{l}})
(M^{2}_{\tilde{l}} - M^{2}_{\tilde{\chi}^{0}_{1}}) }}
{M_{\tilde{l}} }
\end{equation}
As discussed in Ref. \cite{ref1} for \chha \chnb \ Electro-Weak (EW)
production, the three-body leptonic decays of \chnb \ can be used to
measure the mass difference between \chnb \ and \chna.
At LHC, however, much higher production
cross-sections are expected for strongly interacting gluinos
and squarks.
If  \q \ (\g) is lighter than \g \ (\q), it
 can decay to quark plus neutralino or chargino:
\q $\rightarrow q \tilde{\chi}^0_i / q'\tilde{\chi}^{\pm}_i$
(\g $\rightarrow q q' \tilde{\chi}^0_i /
q q'\tilde{\chi}^{\pm}_i$). Eventually,
this source of \chnb \ is much more prolific than direct EW production.
Therefore, we propose to perform inclusive search for \chnb \
and to use the specific shape of the dilepton mass spectrum
as evidence for SUSY \cite{ref2}.
For the SM background suppression,
besides two same-flavour opposite-sign leptons,
one can ask an additional signature characterising SUSY.
This can be: missing energy taken away by escaping 
LSPs, or additional jets coming from gluinos and squarks cascade decays,
or an extra high-$p_T$ lepton from charginos, neutralinos,
sleptons, IVBs, or b-quarks copiously produced in SUSY.
Depending on sparticle masses and predominant decay modes one of these
extra requirements can turn out to be more advantageous than the others,
or to be complementary.

In this paper we discuss
the possibility to observe the \chnb \ in the 
inclusive 3~$leptons$ and 3~$leptons$ + \etm \ final states with the CMS
detector at LHC.
The search strategy for SUSY is to look for a characteristic shape
in the dilepton invariant mass spectrum with a sharp edge, rather than
just for an event excess over the SM expectations. 
Such an observation would reveal SUSY through \chnb \ production,
and at the same time would allow to constrain sparticle
masses and some of the SUSY model parameters.
We also propose a way to resolve the \chnb \ decay type
ambiguity (three-body versus two-body), and a method
of slepton mass determination in the \chnb \ two-body decay
chain.

The present analysis is performed within the framework of
the minimal Supersymmetry Model motivated by Supergravity (mSUGRA),
where only five extra parameters are introduced~\cite{ref3}:
$m_0$, the universal scalar mass,
$m_{1/2}$, the universal gaugino mass and
$A_0$, the universal trilinear term at GUT scale;
$tan \beta$, the ratio of vacuum expectation values of two
Higgs fields;
$sign(\mu)$, the sign of the Higgsino mass parameter.   
Once these parameters are specified all sparticle masses
and couplings at EW scale are evolved via the renormalization group
equations. In the following we limit ourselves to the set:
$tan \beta = 2, \  A_0 = 0, \  \mu < 0$ and investigate
what is the region of ($m_0, m_{1/2}$) parameter space,
where the dilepton invariant mass edge is visible
\cite{ref2}.

\section{$\tilde{\chi}^{0}_{2}$ production and decay}

Within mSUGRA the following relations are valid:

\begin{equation}
M_{\tilde{\chi}^{0}_{2}} \approx
M_{\tilde{\chi}^{\pm}_{1}} \approx 2 M_{\tilde{\chi}^{0}_{1}}
\approx (0.25 \div 0.35) M_{\tilde{g}} \approx 0.9 m_{1/2}
\end{equation}  

\begin{equation}
M_{\tilde{l}_L}^2 =
m_0^2 + 0.52\ m_{1/2}^2 - 0.5(1-2
\mathrm{sin}^{2}\theta_{W}) \ m_Z^2 \ \mathrm{cos}2\beta
\end{equation}
\begin{equation}
M_{\tilde{l}_{R}}^{2} = m_0^2 + 0.15 \ m^2_{1/2} -   
\mathrm{sin}^{2}\theta_{W} \ m_Z^2 \ \mathrm{cos}2\beta
\end{equation}
The three-body leptonic decays of \chnb \ dominate
below $m_{1/2} \sim 200$ GeV nearly for all values of
$m_{0}$, and there is also a small region at higher $m_{1/2}$
where these decays are open.
In these regions 
the measurement of $M^{max}_{l^{+}l^{-}}$ within mSUGRA
allows to get estimate of
$M_{\tilde{\chi}^{0}_{1}}$ (\chna= LSP), $M_{\tilde{\chi}^{0}_{2}}, \
M_{\tilde{\chi}^{\pm}_{1}}, \ M_{\tilde{g}}$
and $m_{1/2}$ (c.f. eqs. (1), (3)).
At $m_{1/2} \sim 200$ GeV, the  modes,
\chnb \ $\rightarrow$ h\chna/Z\chna, become
kinematically allowed thus suppressing direct leptonic decays of
\chnb.
The two-body decays of the \chnb \
via $\tilde{l}_{L/R}$
occur mostly at a low values of $m_{0}$.
In these regions the measurement of only the edge position does not
provide information on masses unambiguously (c.f. eq. (2)). However,
as shown later, by analysing some kinematical 
distributions, the masses of \chna, \chnb \ and of the intermadiate
slepton can be determined with a reasonable precision.
Therefore, with  two-body decays one can constrain
$m_{1/2}$ and $m_0$ if the information on
tan$\beta$ is available, e.g., from the Higgs sector of the
model.
In some regions of parameter space 
the two-body and three-body leptonic decays coexist
or both modes \chnb \ $\rightarrow l^{\pm} \tilde{l}^{\mp}_R$
and \chnb \ $\rightarrow l^{\pm} \tilde{l}^{\mp}_L$ are open,
leading to a possibile observation of a "double edge".

\begin{figure}[hbtp]

\vspace{-10mm}

\hspace*{-0mm}  
\resizebox{13.cm}{!}{\includegraphics{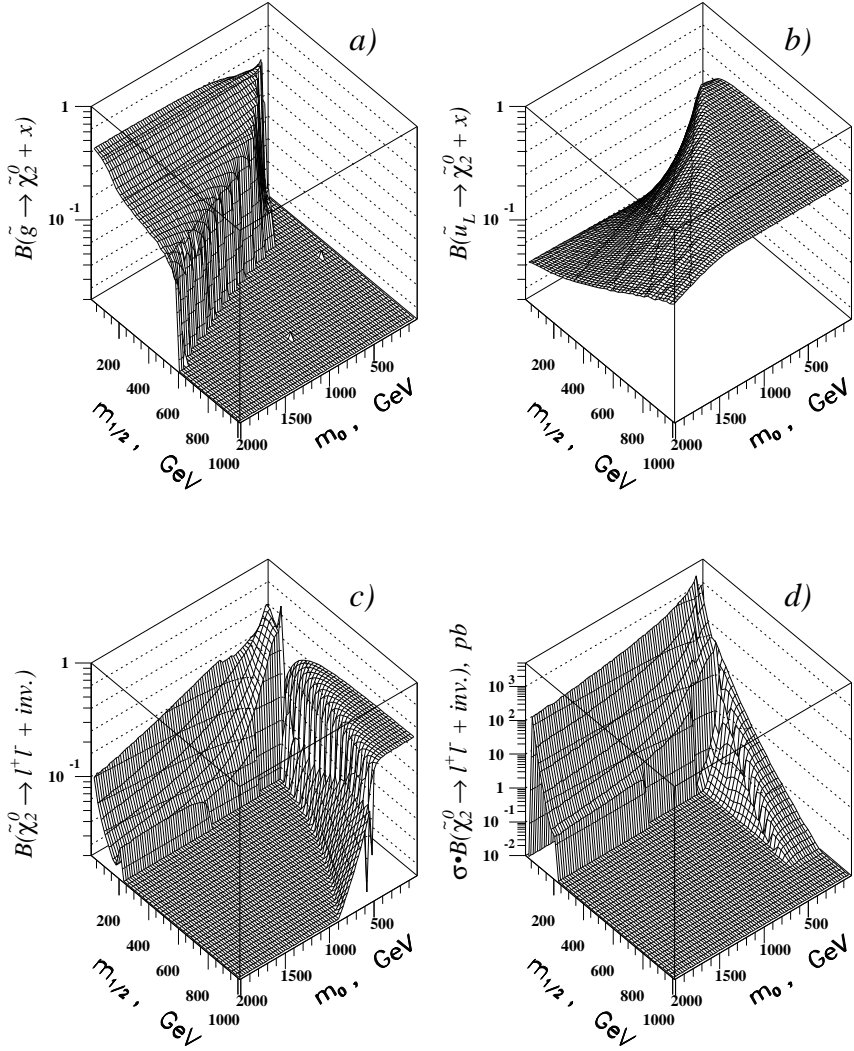}}
 \vspace{-5mm}
   \caption{ \ $\tilde{\chi}^{0}_{2}$ inclusive production
and decay: \ a) \ branching ratio
B($\tilde{g} \rightarrow \tilde{\chi}^{0}_{2} +$ x), \
b) \ B($\tilde{u}_{L} \rightarrow \tilde{\chi}^{0}_{2} +$ x), \  
c) \ B($\tilde{\chi}^{0}_{2} \rightarrow l^+l^- + invisible$) \
and \ d) \ $\tilde{\chi}^{0}_{2}$ inclusive production cross-section
times branching ratio into $l^+l^-$ ($l = \mu$ or e).}
\end{figure}

At LHC the \chnb \ is abundantly produced 
either from gluinos or from
squarks over almost the whole ($m_{0}, m_{1/2}$) plane,
as illustrated in Figs.$1a, b$. Figure $1c$ shows the \chnb \ decay
branching ratio into leptons ($l = \mu$ or e). Finally,
the \chnb \ inclusive production cross-section times
branching ratio into leptons is shown in Fig.$1d$.
Clearly, one expects quite large rates of
opposite-sign same-flavour dileptons from inclusive \chnb \ production
over a large portion of ($m_{0}, m_{1/2}$).
Similarly to the \chnb, the lightest chargino is also copiously 
produced from strongly interacting sparticles.
Moreover, the leptonic decay branching of the \chha \
always exceeds 0.1 per lepton flavour.
For low values of $m_0$, in the regions where decays to sleptons are
kinematically allowed, the \chha \ gives a lepton with a probability
close to 1. Making use of this 
prolific production of leptons from \chha, as well as
from other SUSY sources,
the SM background can be suppressed by asking for an
additional high-$p_T$ lepton.
Thus the \chnb \ inclusive production can be studied in
3~$lepton$ and 3~$lepton$ + \etm \ final states.

\section{Event simulation and selection}

For signal simulation we have generated
all mSUGRA processes, using ISAJET 7.14 \cite{ref4}.
The considered SM sources of background,
WZ, $t \bar{t}$, ZZ and Z$b \bar{b}$ productions,
have been  simulated by P{\footnotesize YTHIA} 5.7 \cite{ref5}.
The following production cross-sections have been assumed at
LHC energy of $\sqrt{s}=14$ TeV:
$\sigma_{WZ} = 26$ pb, $\sigma_{t \bar{t}} = 670$ pb
for a top mass of 170 GeV, $\sigma_{ZZ} = 15$ pb and
$\sigma_{Z b \bar{b}} = 580$ pb.

We have used a parameterised detector simulation
program \cite{ref6}. It takes into account: i) particle bending in
a 4 T magnetic field; ii) smearing of lepton
momentum according to parameterisation obtained from detailed
simulations; iii) $90\%$
triggering plus reconstruction efficiency
per lepton within geometrical acceptance 
$|\eta| < 2.4$; iv) $90 \%$ reconstruction efficiency  per charged track
with $p_T > 1$ GeV within $|\eta| < 2.4$;
v) coverage, granularity, main cracks, energy resolution  
and electronic noise of the calorimetry system;
vi) fluctuation of starting point and the
spatial development of electromagnetic and hadronic showers
by parameterisation of the lateral and longitudinal profiles.
For the high luminosity study, $L=10^{34}$
cm$^{-2}$s$^{-1}$, on top of each signal and background events
we have superimposed on average 15 "hard"
pile-up events (P{\footnotesize YTHIA} QCD jet production with
$\hat{p}_T>5$ GeV) fluctuated by Poissonian distribution.

In the inclusive 3~$lepton$ channel we require: 
1) two opposite-sign, same-flavour leptons ($\mu$ or e) with 
$p_T > 10$ GeV;
2) a third "tagging" lepton with $p_T > 15$ GeV;
3) lepton isolation: if there is  no track with $p_T > 2 \ (1.5)$
GeV within
$\Delta R = \sqrt{\Delta \eta^2 + \Delta \phi^2} < 0.3$
around the lepton direction, it is considered as isolated.
The leptons $p_T$ thresholds are variable
depending on the region of ($m_0, m_{1/2}$) studied.
In some cases leptons are not required to be
isolated.
In the inclusive 3~$lepton$ + \etm channel, beside the above three
requirements we also ask: 4) \etm $>$ 200 (300) GeV.

In the selected events we reconstruct
the invariant mass $M_{l^{+}l^{-}}$ of the lepton pair
with the same flavour and opposite sign. 
When several $l^{+}l^{-}$ combinations per event are possible
(in case of, e.g., three leptons of the same flavour) the one
with the minimal distance $\Delta R^{l^{+}l^{-}}$ is chosen.

\section{Dilepton invariant mass spectrum}

\begin{figure}[hbtp]

\vspace{-13mm}

\hspace*{-5mm}
\resizebox{13.5cm}{!}{\includegraphics{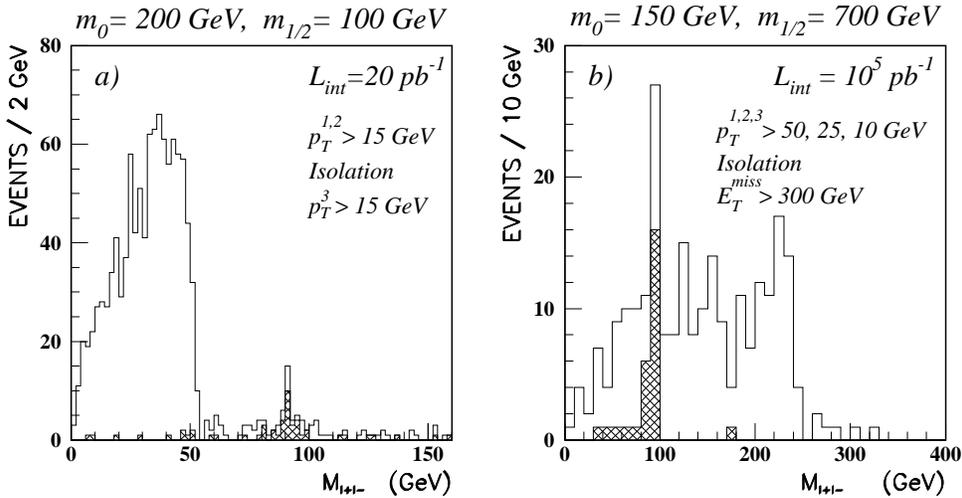}}
   \caption{Expected $l^+l^-$ mass spectrum for mSUGRA points:
a) $m_0 =200$ GeV, $m_{1/2} =100$ GeV in the inclusive 3~$lepton$
final states;
b) $m_0 =150$ GeV, $m_{1/2} =700$ GeV in the 3~$lepton$ + \etm \
final states.
The other mSUGRA parameters are tan$\beta$=2, $A_{0} =0$ and $\mu
< 0$. The hatched histogram corresponds to the SM background.}
\end{figure}

The dilepton invariant mass spectrum for a representative
mSUGRA parameter space point ($m_0=$ 200 GeV, $m_{1/2}$ = 100
GeV) superimposed on the expected SM background is shown in Fig.$2a$. The
number of events corresponds to an integrated luminosity of 
$L_{int}=20$ pb$^{-1}$, i.e. a few hours of LHC running
at a luminosity of $L=10^{33}$ cm$^{-2}$s$^{-1}$.
The $p_T$ thresholds on all three leptons are 15 GeV 
and two of them entering
mass distribution are isolated.
The SM background is
negligible, with the dominant contribution from WZ production
around the Z mass peak and
$t \bar{t}$ production at lower invariant masses.
The specific shape of the distribution with its sharp
edge reveals SUSY through \chnb \ production.
At this mSUGRA point the \chnb \ has three-body decay modes,
and  the  relevant masses are $M_{\tilde{\chi}^{0}_{2}}
= 97$ GeV, $M_{\tilde{\chi}^{0}_{1}} = 45$ GeV.
The edge is situated at the expected value of
$M^{max}_{l^{+}l^{-}} = M_{\tilde{\chi}^{0}_{2}} -
M_{\tilde{\chi}^{0}_{1}} = 52$ GeV and its position can be measured
with an accuracy better than 0.5 GeV.
Within the model this allows to determine
$M_{\tilde{\chi}^{0}_{1}}, \ M_{\tilde{\chi}^{0}_{2}}, \
M_{\tilde{\chi}^{\pm}_{1}}, \ M_{\tilde{g}}$
and $m_{1/2}$ (c.f. (3)). 
Note, that
the Z peak seen in Fig.2 serves as an overall calibration
signal; it allows to control the mass scale as well as the production 
cross-section.

With increasing $m_{1/2}$ the observation of the "edge" becomes 
more difficult due to a rapid fall of gluino/squark production
cross-sections. Here, on the other hand, heavier
\g \ and \q \ provide higher
missing transverse energy. Asking \etm $>$ 200 (300) GeV
rejects most of the SM background whilst retaining a large fraction of
signal
events. Figure $2b$ shows the dilepton spectrum for the mSUGRA point
(150 GeV, 700 GeV)
in the 3~$lepton$ + \etm \ final states. In this region of high $m_{1/2}$
dileptons
are mainly produced through cascade decays of \chnb \ through
intermediate state sleptons. The third,
"tagging" lepton, predominantly comes from a \chha \
produced in association, again from $\tilde{g}, \tilde{q}$ cascades.

\begin{figure}[hbtp]

\vspace{-13mm}

\hspace*{5mm}  
\resizebox{12.cm}{!}{\includegraphics{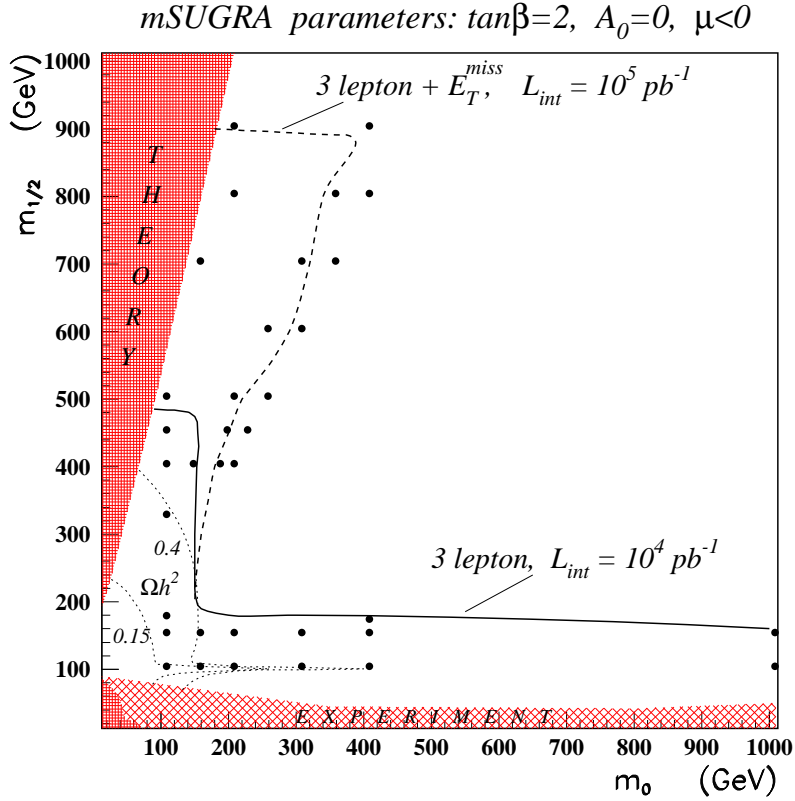}}

\vspace{-13mm}

   \caption{Explorable region of mSUGRA in inclusive
3~$lepton$ and
3~$lepton$ + \etm \ final states with
$L_{int}= 10^4$ pb$^{-1}$ and
$L_{int}= 10^5$ pb$^{-1}$, respectively.
Shaded regions are excluded by theory
and/or experiment. 
The simulated mSUGRA points are shown by circles.
The cosmologically preferred
region 0.15 $< \Omega h^2<0.4$ is also given.}

\vspace{-0mm}

\end{figure}

In order to delineate the region of parameter space
where the $l^+l^-$ edge is visible, the $(m_{0}, m_{1/2})$ plane
has been systematically scanned for fixed
tan$\beta$ = 2, $A_0 = 0$ and $\mu < 0$.
At least 40 signal events in a mass interval $>10$ GeV
just below the edge position and a statistical significance
$S = N_S / \sqrt{N_S + N_B} > 7$ are required
($N_S$ and $N_B$ are the number of signal and background events in this
mass interval). Figure 3 shows
the domain explorable under these conditions 
with an integrated luminosities
of $L_{int} = 10^{4}$ pb$^{-1}$ and $L_{int} = 10^{5}$ pb$^{-1}$
for 3~$lepton$ and 3~$lepton$ + \etm \
channels, respectively.
Of particular importance is the fact, that
already at $L_{int} = 10^{4}$ pb$^{-1}$   
the cosmologically preferred region
$0.15 \lsim \Omega h^2 \lsim 0.4$ \cite{ref7}, where
the \chna \ would be a good candidate for dark matter,
is entirely covered; here $\Omega$ is the
ratio of the \chna \ relic density to the critical density of the
Universe and $h$ is the Hubble constant scaling factor.
In the 3~$lepton$ + \etm \ final states with $L_{int} = 10^5$ pb$^{-1}$
the reach extends up to $m_{1/2} \sim 900$.
A detectable edge is seen as
long as $\sigma \cdot $B $\gsim 10^{-2}$ pb (see Fig.$1d$).

\section{\chnb \ decay type and slepton mass determination}

\begin{figure}[hbtp]

\vspace{-10mm}

\begin{minipage}[b]{.48\linewidth}
\hspace*{-5mm}
\resizebox{7.cm}{!}{\includegraphics{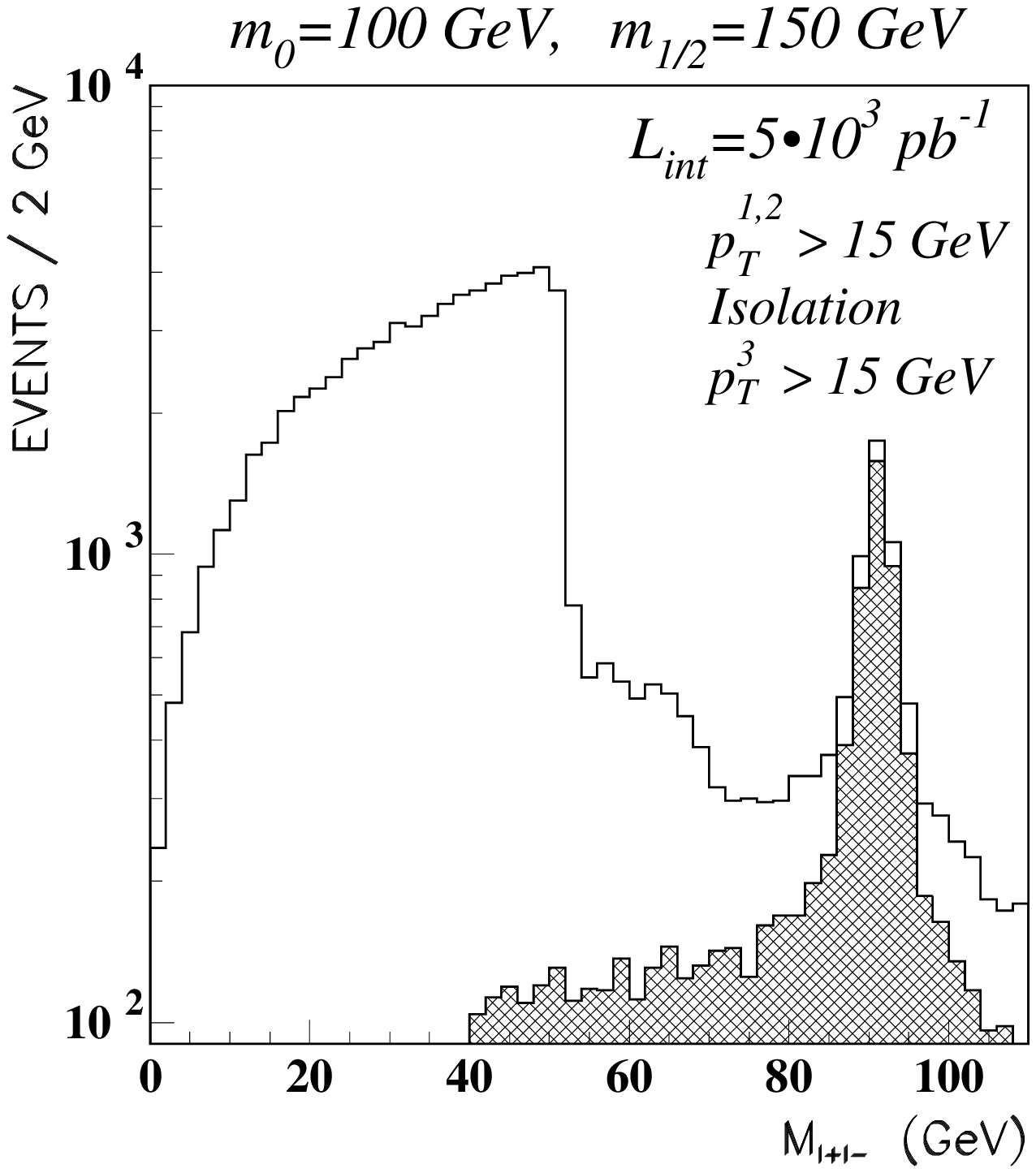}}
\vspace{-13mm}
 \caption{ $l^+l^-$ mass distribution for $m_0$ =
100 GeV, $m_{1/2} =150$ GeV,
 tan$\beta$=2, $A_{0} = 0$ and $\mu < 0$.
The hatched histogram corresponds to the SM background. }
\end{minipage}\hfill
\begin{minipage}[b]{.48\linewidth}
\hspace*{-5mm}
\resizebox{7.cm}{!}{\includegraphics{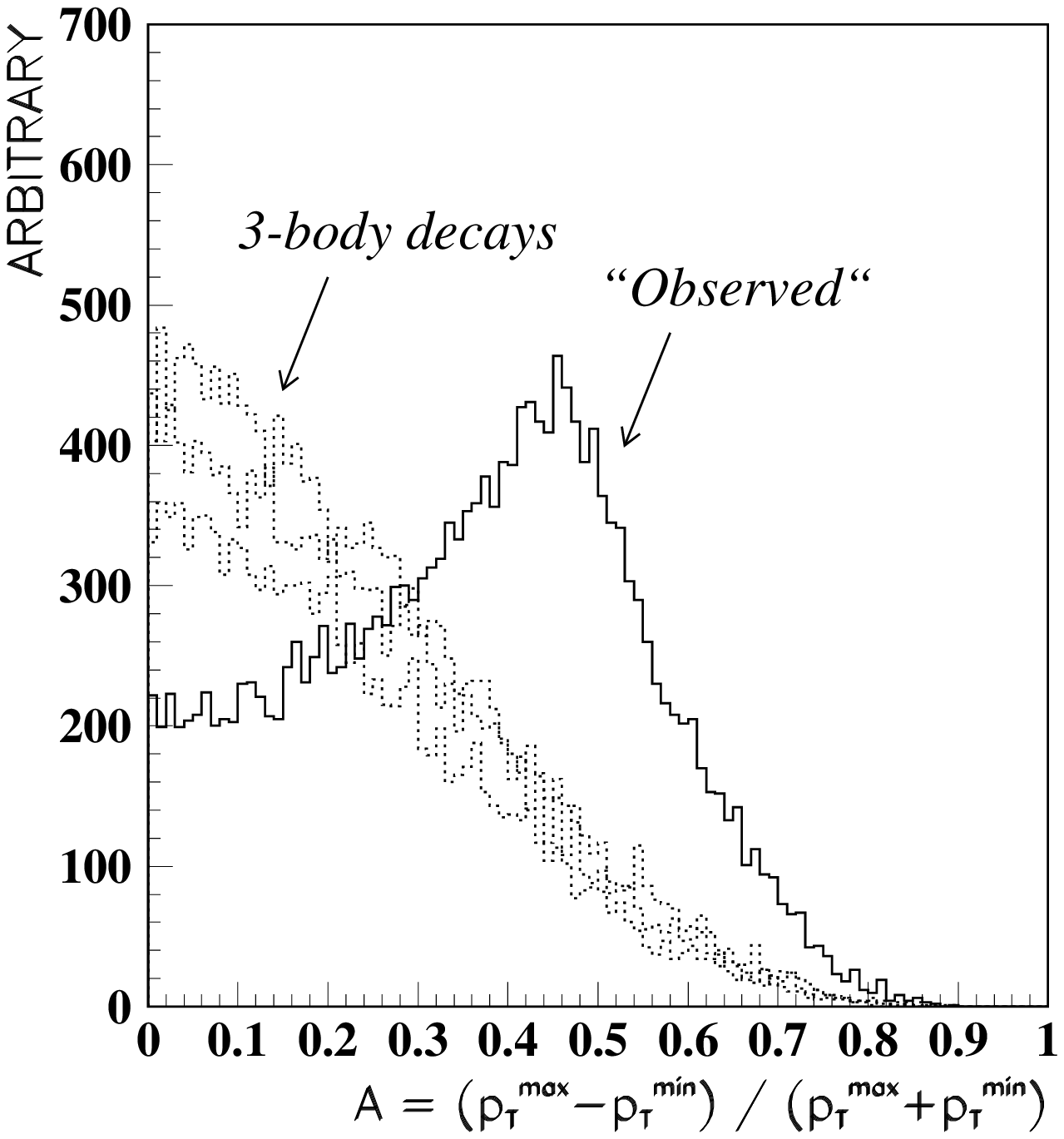}}
\vspace{-13mm}
   \caption{Lepton $p_T$-asymmetry distributions in
three- and two-body decays of \chnb \ with the
same position of dilepton invariant mass
edge.}
\vspace{4mm}
\end{minipage}
\end{figure}

The \chnb \ decay type (three-body versus  two-body)
determination is essential
for the interpretation of the observed edge.
The $p_T$ spectra of leptons from \chnb \ (cascade) two-body decays are
more asymmetric compared to the three-body ones.
To characterise this asymmetry
we introduce the variable  
$\mathcal{A} = \frac{p_T^{max} - p_T^{min}} {p_T^{max} + p_T^{min}}$,
where $p_T^{max}$ ($p_T^{min}$) corresponds to the lepton
of maximal (minimal) transverse momentum. 
An example of decay-type determination is given for the
mSUGRA point (100 GeV, 150 GeV). 
Here the \chnb \ decays via
$\tilde{\chi}^{0}_{2} \rightarrow l^{\pm} \tilde{l}_{R}^{\mp}
\rightarrow
l^+l^- \tilde{\chi}^{0}_{1}$
with 0.54 branching ratio and
the sparticle masses are
$M_{\tilde{\chi}^{0}_{2}} = 135$ GeV,
$M_{\tilde{\chi}^{0}_{1}} = 65$ GeV,
$M_{\tilde{l}_{R}} = 120$ GeV.
Figure 4 shows the dilepton spectrum
for this point with 
$L_{int} = 5 \cdot 10^3$ pb$^{-1}$,
superimposed on the SM background, in the inclusive 3~$lepton$
final states. 
An edge is situated at $\sim$ 52 GeV, as expected.
To look at the lepton $p_T$-asymmetry distribution,
we pick up the lepton pairs with $M_{l^+l^-} < 52$ GeV.
Figure 5 (full line) shows the corresponding $\mathcal{A}$ spectrum
for the selected events. The pronounced asymmetry in $p_T$
indicates the cascade nature of \chnb \ decays.
The dotted histograms are obtained with the 
assumption that the "observed" edge is due to direct three-body decays
with various $M_{\tilde{\chi}_2^0}$, $M_{\tilde{\chi}_1^0}$ combinations
providing $M_{\tilde{\chi}_2^0} - M_{\tilde{\chi}_1^0} = 52$ GeV.
In these cases the asymmetry $\mathcal{A}$ distributions peak at zero.

After the decay type is determined
a further analysis step is carried out to extract the masses
of involved particles:
i) assume $M_{\tilde{\chi}^{0}_{2}} = 2 M_{\tilde{\chi}^{0}_{1}}$,
ii) generate samples of $\tilde{\chi}^{0}_{2}$ 2-body cascade
decays for various $M_{\tilde{\chi}^{0}_{1}}$
($M_{\tilde{l}}$); note, that the slepton mass is constrained
to provide the "observed" position of an edge given by eq. (2),
iii) the best combination of $M_{\tilde{\chi}^{0}_{1}}$,
$M_{\tilde{\chi}^{0}_{2}}$ and $M_{\tilde{l}}$
is then chosen by means of a $\chi^2$-test of
the shape of the lepton $p_T$-asymmetry distributions.
This procedure provides the following precisions on masses:
$\delta M_{\tilde{\chi}^{0}_{1}} \lsim$ 5 GeV,
$\delta M_{\tilde{l}} \lsim$ 10 GeV.
The use of
the total number of observed events, as well as the dilepton
invariant mass spectrum itself in a combined
$\chi^2$-test would further improve these results.

\section{Double edge}

\begin{figure}[hbtp]

\vspace{-10mm}

\begin{minipage}[b]{.48\linewidth}
\hspace*{-5mm}
\resizebox{7.cm}{!}{\includegraphics{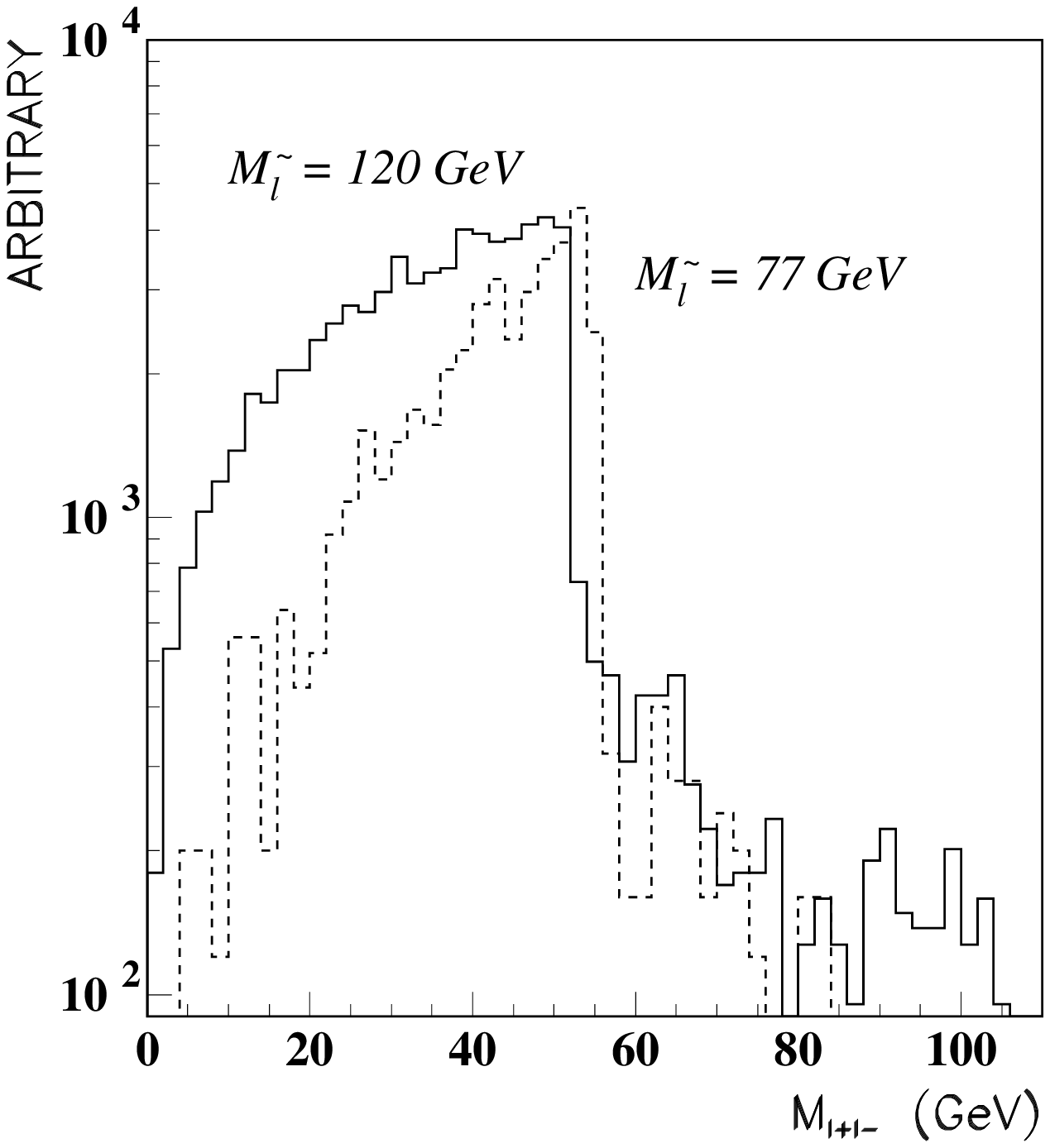}}
\vspace{-13mm}
   \caption{Predicted $l^+l^-$ mass spectra for
two values of slepton mass: $M_{\tilde{l}}=120$ GeV (full line)
and $M_{\tilde{l}}=77$ GeV (dashed line); mSUGRA parameters
are as in Fig.4.}
\end{minipage}\hfill
\begin{minipage}[b]{.48\linewidth}
\hspace*{-5mm}
\resizebox{7.cm}{!}{\includegraphics{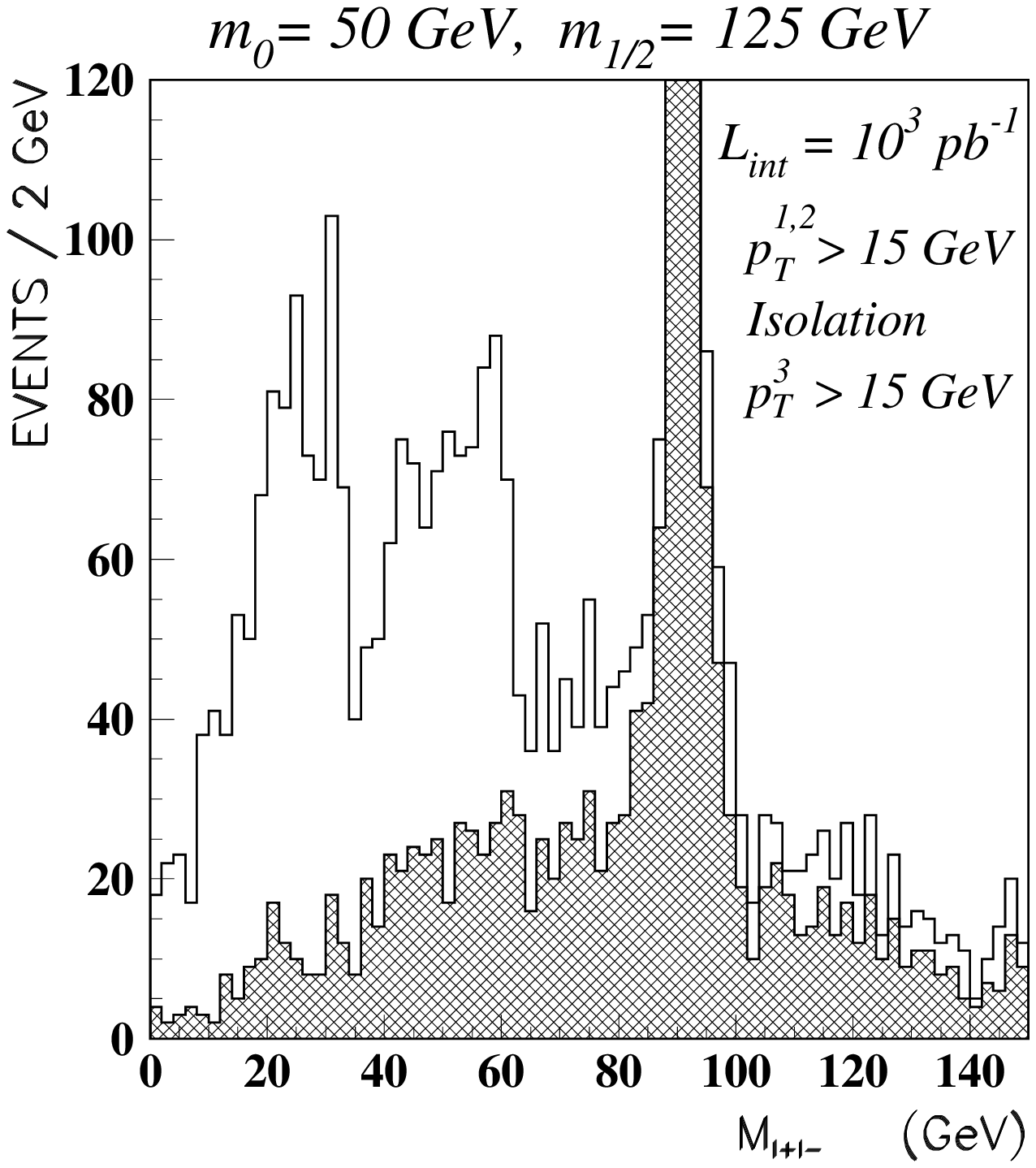}}
 \vspace{-13mm}
   \caption{Expected $l^+l^-$ mass spectrum for mSUGRA Point
 $m_0 =50$ GeV, $m_{1/2} =125$ GeV, tan$\beta$=2, $A_{0} =0$ and
$\mu <
0$. The hatched histogram corresponds to the SM background.}
\end{minipage}\hfill
\end{figure}

As discussed in section 2, in some regions of the
($m_0, m_{1/2}$) plane the different leptonic decay modes of \chnb \
can be simultaneously open and
observation of "double edges" is possible if the branching ratios are
not too dissimilar. This is the case, for instance, for the mSUGRA point
(100 GeV, 150 GeV) discussed in the previous section.
Here, beside the
$\tilde{\chi}^{0}_{2} \rightarrow l^{\pm} \tilde{l}_{R}^{\mp}
\rightarrow l^+ l^- \tilde{\chi}^{0}_{1}$ decays, the
direct three-body decay mode is also open, though,
with a much smaller branching ratio,
B$(\tilde{\chi}^{0}_{2} \rightarrow l^{+}l^{-}
\tilde{\chi}^{0}_{1})=0.05$.
The second edge 
at $M^{max}_{l^{+}l^{-}}=69$ GeV seen in Fig.4 is
thus less spectacular than the one at 52 GeV.
The $p_T$-asymmetry distribution
of dileptons with 55 GeV $< M_{l^+l^-} <$ 69 GeV
indicates the three-body decay type of these events.
Using now two measured values of edge positions and
assuming
$M_{\tilde{\chi}^{0}_{2}} = 2 M_{\tilde{\chi}^{0}_{1}}$,
we obtain two solutions for the slepton mass
from equations (1) and (2):
$M_{\tilde{l}} = 120$ GeV  and  77 GeV.
Corresponding dilepton invariant mass spectra for these two
solutions are shown in Fig.6. The clear difference in the predicted
spectra (Fig.6 vs. Fig.4) allows to eliminate the $M_{\tilde{l}} = 77$ GeV
solution.
At this particular mSUGRA point the use of both edges
provides a precision of
$\delta M_{\tilde{\chi}^{0}_{1},\tilde{l}} \lsim$ 1 GeV.

Another example of a "double edge" is given in Fig.7
for the mSUGRA point (50 GeV, 125 GeV). Here
the sparticle masses are
$M_{\tilde{\chi}^{0}_{2}} = 116$ GeV,
$M_{\tilde{\chi}^{0}_{1}} = 55$ GeV,  
$M_{\tilde{l}_{L}} = 110$ GeV,
$M_{\tilde{l}_{R}} = 78$ GeV and now two
two-body decays coexist, with a comparable branching ratios
B($\tilde{\chi}^{0}_{2} \rightarrow l^{\pm} \tilde{l}^{\mp}_{L}
\rightarrow l^+  l^- \tilde{\chi}^{0}_{1}$) = 0.037 and 
B($\tilde{\chi}^{0}_{2} \rightarrow l^{\pm} \tilde{l}^{\mp}_{R}
\rightarrow l^+  l^- \tilde{\chi}^{0}_{1}$) = 0.013
and their analysis could proceed as indicated above
providing a strong constraint on the underlying model.

\section{Conclusions}

$\bullet$
Observation of an edge in the dilepton invariant mass spectrum
reflects production of $\tilde{\chi}^{0}_{2}$ and hence
establishes existence of SUSY; this observation
in some cases is possible with very small statistics and could be
the first evidence for SUSY at LHC.

$\bullet$
In the inclusive 3~$lepton$ final states, with
no great demand on detector performance, a large portion of parameter
space, including the cosmologically
preferred mSUGRA domain, can be investigated at an
integrated luminosity of $L_{int} = 10^{4}$ pb$^{-1}$.

$\bullet$
In the inclusive 3~$lepton$ + \etm \ final states and with
$L_{int} = 10^{5}$ pb$^{-1}$ most of the region where the
$\tilde{\chi}^{0}_{2}$ inclusive production cross-section times
branching ratio into leptons $\gsim 10^{-2}$ pb can be covered.

$\bullet$ From the observation of characteristic
edge(s) and with mSUGRA motivated assumptions,
it is possible to determine masses of
$\tilde{\chi}^{0}_{1}$ and $\tilde{\chi}^{0}_{2}$
in case of direct decays
and masses of 
$\tilde{\chi}^{0}_{1}$, $\tilde{\chi}^{0}_{2}$
and sleptons in case of cascade decays.

\vspace{2mm}

{\large {\bf {Acknowledgements}}}

\vspace{0mm}

We would like to thank Daniel Denegri for many
fruitful discussions and encouragement, Walter Geist, Tejinder Virdee and
John Womersley for useful remarks.

\end{document}